# Crystal structural evolution of Ru$_3$Sn$_7$ under pressure and its implication on possible electronic changes

K. A. Irshad[1*], P. Anees[2], Hrudananda Jena[2] and Boby Joseph[1†]

[1]Elettra - Sincrotrone Trieste S.C.p.A. S.S. 14 Km 163,5 in Area Science Park
34149 Basovizza, Trieste, Italy
[2]Indira Gandhi Centre for Atomic Research, A CI of Homi Bhabha National Institute,
Kalpakkam, 603102, Tamilnadu, India

**ABSTRACT**. Ru$_3$Sn$_7$, an intermetallic compound with advanced catalytic properties, exhibits a complex crystal structure and intriguing electronic properties, making it an attractive candidate for investigations under high-pressure (HP). The structural, vibrational and electronic band structure of this compound were investigated at HP up to ~ 20 GPa using synchrotron x-ray powder diffraction, micro-Raman, and density functional theory (DFT), respectively. Despite the local structural changes implied by a discernible reduction in the compressibility and distinct slope changes in the pressure evolution of the symmetric stretching vibrations of the Ru and Sn atoms around 8 GPa, the cubic structure is found to be stable throughout the pressure range. In support, our calculated phonon dispersion relation confirmed the stability of the cubic phase till the highest pressures. A comprehensive analysis of the Raman spectrum reveals the signatures of the pressure induced sudden strengthening of electron-phonon coupling as early as 3 GPa which is backed by a bounce in the phonon and electron density of states (DoS). Electronic structure calculations demonstrate that the metallic nature of Ru$_3$Sn$_7$ is preserved in the studied pressure range with a minor redistribution of electronic DoS across the Fermi level (E$_F$). The band structure calculations predict intriguing changes in the electronic structure, revealing the pressure induced *dp* hybridization through the high symmetry point of the Brillouin zone which is largely responsible for the observed reduction in the compressibility and enhancement of the electron-phonon coupling in Ru$_3$Sn$_7$.

*Contact author: irshad.kariyattuparamb@elettra.eu
†Contact author: boby.joseph@elettra.eu

**I. INTRODUCTION.**

Intermetallic Ru$_3$Sn$_7$ has shown its potential as a catalyst in chemical reactions, an electrode material in energy storage devices, and a semiconductor in electronic applications [1-6]. The intriguing electronic properties of this material can be closely linked to its complex crystal structure [1,7-10]. Ru$_3$Sn$_7$ belongs to the family of Ir$_3$Ge$_7$ intermetallics crystallizes in the cubic structure with $Im\bar{3}m$ space group [9,10]. In the cubic lattice, Ru is bonded to eight Sn atoms in square antiprismatic geometry with four shorter and four slightly longer Ru-Sn bonds. In the two inequivalent Sn sites, one of them is bonded to four equivalent Ru atoms in a 4-coordination geometry while the other Sn is bonded to three equivalent Ru and four equivalent Sn atoms forming a 7-coordination environment (see the inset of Fig. 1). The detailed crystal structure is explained in a later section and elsewhere in the literature [9,10]. Compounds like Ir$_3$Sn$_7$, Mo$_3$Sb$_7$ Ni$_3$Ga$_7$, Pd$_3$In$_7$, Os$_3$Sn$_7$, and Re$_3$As$_7$, are some of the isostructural compounds that are identified in this family of intermetallic compounds [11-13]. Among those, Mo$_3$Sb$_7$ has been extensively investigated due to its thermo-electric and superconducting nature [14-18]. Though the superconductivity was absent, Ru$_3$Sn$_7$ and some other related compounds have attracted considerable attention mainly due to their catalytic and thermoelectric properties [1,2,19,20].

Limited reports are available on the electronic structure of Ru$_3$Sn$_7$ and its family of isostructural compounds in the literature. Density functional theory (DFT) calculations have shown a continuous distribution of the electronic density of states (DoS) across the Fermi energy ($E_F$) leading to the conclusion that Ru$_3$Sn$_7$ is metallic in nature [8,21]. However, recent electronic structure calculations have revealed the existence of nontrivial topological surface states in Ru$_3$Sn$_7$ [1]. These surface states have been beneficial in the hydrogen evolution reaction due to the increased carrier mobility. Recent transport measurements coupled with the electronic structure calculations on the isostructural $Pd_3In_7$ have revealed the Dirac type-II metallic nature in this system [13]. Similar Dirac-like dispersion has also been reported in the isostructural $Rh_3In_{3.4}Ge_{3.6}$ where the electronic structure contains several Dirac type-I, type-II, and type-III nodes [8]. In contrast to the metallic nature of the family, $Mo_3Sb_7$ has been shown to have a bandgap in its electronic structure indicating non-metallic characteristics. The pressure and temperature effect on $Mo_3Sb_7$ has been studied extensively due to the presence of superconductivity in this system. Though there are no structural phase transition discovered at room temperature and higher pressures, the cubic to tetragonal transition in $Mo_3Sb_7$ at 53 K was suppressed by the application of pressure (~10 GPa) [15,17,22]. This transition and the superconductivity was also suppressed by doping of Ru into $Mo_3Sb_7$, possibly due to the modifications in the electronic structure [23]. This is in line with the low-temperature studies on $Ru_3Sn_7$ where no superconductivity was observed down to 2 K [9]. A recent report shows that the substitution of Sb with Sn induces superconductivity in $Ru_3Sn_7$ with an onset transition temperature of 4.2 K [24]. The substitution of Sb with an additional 5$p$ electron introduces more homogenous free electrons in $Ru_3Sb_{1.75}Sn_{5.25}$ leading to the emergence of superconductivity. As pressure is a clean external control parameter, which can systematically tune the structure, high-pressure (HP) studies are highly desirable to better understand and optimize electronic properties which are controlling the functional characteristics. While extensive research has been conducted on ambient-pressure behaviour of $Ru_3Sn_7$, as outlined above, there is hardly any data available on its response to HP. Our objective in this work is to explore the HP behavior of $Ru_3Sn_7$ by combining the results obtained from the X-ray diffraction, Raman spectroscopy, and *ab initio* DFT calculations. Our findings are expected to provide insights into pressure evolution of the crystal structure, vibrational and electronic properties of $Ru_3Sn_7$ and its family of intermetallic compounds.

*Contact author: irshad.kariyattuparamb@elettra.eu

†Contact author: boby.joseph@elettra.eu

*Contact author: irshad.kariyattuparamb@elettra.eu
†Contact author: boby.joseph@elettra.eu

## II. EXPERIMENT & COMPUTATION METHODS

The $Ru_3Sn_7$ intermetallic compound was obtained as a by-product during the separation of Ru from waste-loaded borosilicate glass using Sn as the collecting or solvent metal [25]. The alloy button obtained was rich in Sn and $RuO_2$ phases. This was ground well and dissolved in $HNO_3$ to dissolve the excess Sn present in the alloy button. Further, the obtained powder was dissolved in the HCl to dissolve the $RuO_2$/$SnO_2$ present in the alloy button. The powder thus obtained was washed several times with de-ionized water. The phase purity of the sample was ensured in the laboratory diffractometer before the synchrotron diffraction studies.

HP diffraction measurements were carried out at the Xpress beamline of the Elettra Sincrotrone Trieste (Italy). A monochromatic wavelength of 0.4956 Å was used to collect the diffraction data. A BETSA gas-controlled membrane diamond anvil cell (DAC) was used to generate the pressure. Sample, pressure transmitting medium (methanol ethanol solution in the ratio 4:1), and a ruby sphere (pressure calibrant) have been loaded into the hole of 150 μm diameter drilled at the center of the pre-intended steel gasket. The diffraction data were recorded in transmission mode using a PILATUS3S-6M detector. The two-dimensional diffraction data was converted to the one dimensional 2θ Vs intensity data using the Dioptas software [26]. The pressure was calculated from the R1 line shift of the ruby fluorescent spectrum. Crystal structural parameters were obtained by the Rietveld refinement of the diffraction data using the GSAS II programme [27]. Standard $CeO_2$ data collected at a similar configuration was used to obtain the instrument profile function. A customized Renishaw confocal micro-Raman spectrometer equipped with a 532 nm (green) diode laser and a grating of 2400 L/mm, available at the Xpress beamline, was used to acquire the Raman spectrum. Sample assembly and pressure calibration were similar to that of diffraction measurements. Due to the inherent weak nature of the Raman signal, particularly under HP, data was accumulated forty times (with 5 seconds of collection time per accumulation) using a 50x objective at 50 % of the laser power. Under these conditions, we did not observe any laser damage to the $Ru_3Sn_7$ sample throughout HP Raman measurement.

*Contact author: irshad.kariyattuparamb@elettra.eu

†Contact author: boby.joseph@elettra.eu

To understand the electronic structure changes with pressure, spin-polarized density functional theory (DFT) calculations are performed using the Vienna ab initio simulation package (VASP) with projector augmented wave (PAW) pseudo-potentials [28]. Generalized gradient approximation (GGA) of Perdew–Burke–Ernzerhof (PBE) is employed to compute the exchange correlation functional [29]. GGA is found to under-estimate the equilibrium lattice parameter of $Ru_3Sn_7$. However, with the addition of van der Waals (vdW) interaction (incorporated using Grimme's method (DFT-D3) [30]) provided a better agreement to the equilibrium lattice parameters. A kinetic energy cut-off of 520 eV is used to truncate the plane wave basis set. A Monkhorst–Pack [31] grid of size 5x5x5 for structural relaxation and a denser grid of 13x13x13 were used in density of states (DoS) calculations. To explore the topological features, the band structures were computed with and without spin-orbit coupling (SOC) [32]. Density functional perturbation theory (DFPT) [33] calculations are done in conjunction with *phonopy* [34] to study the lattice dynamics properties and then to obtain the Raman frequencies at different pressures.

### III. RESULTS AND DISCUSSION
#### A. Ambient characterization

The X-ray diffraction pattern recorded at ambient pressure and room temperature for the powder sample is shown in Fig. 1. The diffraction pattern shows well-resolved reflections that could be indexed to the known cubic structure (space group: $Im\bar{3}m$) of $Ru_3Sn_7$. No impurity peak was observed, and the Rietveld structure refinement of the diffraction data indicates the single-phase nature of the sample. The lattice parameter, atomic positions, crystallite size, and six background parameters were refined. Isothermal parameters were refined independently. The results of the refinement are shown in Table I.

*Contact author: irshad.kariyattuparamb@elettra.eu

†Contact author: boby.joseph@elettra.eu

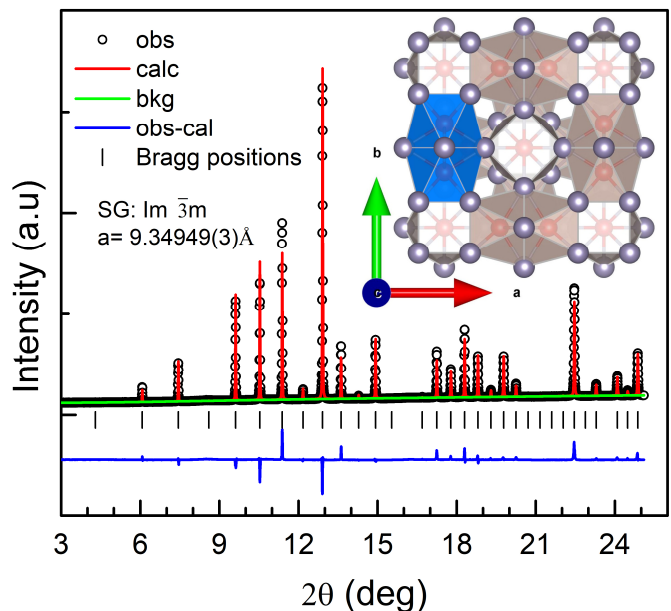

FIG 1. The Rietveld refinement of the X-ray diffraction pattern recorded at the ambient conditions for the $Ru_3Sn_7$. Symbols are data and vertical lines are the Bragg peak positions. Model fit results, background and difference curves are shown in red, green and blue curves, respectively. The inset shows the polyhedral representation of the crystal structure of $Ru_3Sn_7$. Ru atoms are indicated in red color. The two Ru-Sn polyhedrons with square antiprism arrangement are connected through their square face (involving only Sn(1) atoms) and are represented with a blue-coloured polyhedron

The crystal structure of $Ru_3Sn_7$ belongs to the same family as that of $Ir_3Ge_7$ and $Mo_3Sb_7$. A polyhedral representation of the crystal structure of $Ru_3Sn_7$ is shown in the inset of Fig. 1. In the $Ru_3Sn_7$, the Ru atom occupies the 12$e$ Wyckoff position whereas the Sn atom occupies two different positions, Sn(1) at 12$d$ and Sn(2) at 16$f$. These Sn atoms form a square antiprism arrangement around the central Ru atom in which the Sn(1) and Sn(2) independently occupy the two square faces. Face sharing of two such polyhedral units through the square face occupied by Sn(1) results in the building block of the $Ru_3Sn_7$ framework. These dimer structures are connected through the Sn(2)-Sn(2) edges to form the framework structure of $Ru_3Sn_7$. Two such frameworks were interconnected through the corner sharing of the Sn(1) atoms to form the crystal structure of Ru$_3$Sn$_7$. Table II shows a comparison of the lattice parameter obtained from our experiment and DFT calculations. The GGA functional under-estimate the lattice parameter, when compared to experiments by ~ 4.57 %. However, the incorporation of the Van der Waals (vdW) interaction resulted in very good agreement (overestimated only by 0.25 %) with the experimental value of 9.34349(3) Å. This suggests that the incorporation of the vdW interactions is necessary to reproduce the equilibrium properties.

*Contact author: irshad.kariyattuparamb@elettra.eu
†Contact author: boby.joseph@elettra.eu

TABLE 1. The Rietveld refined crystal structure of Ru$_3$Sn$_7$ at ambient conditions. A lattice parameter of $a$ = 9.34949(3) Å is obtained using the space group $Im\bar{3}m$ (n. 229) during the refinement. The isothermal atomic displacement parameter (ADP), U$_{iso}$ = 0.0084 Å$^2$.

| Atom (WP) | Fractional atom positions | | |
|---|---|---|---|
| | x | y | z |
| Ru(12$e$) | 0.34740(28) | 0 | 0 |
| Sn(12$d$) | 0 | 0.5 | 0.75 |
| Sn(16$f$) | 0.16179(12) | 0.16179(12) | 0.16179(12) |

WP is the Wyckoff position

TABLE 2. The comparison of the lattice parameter obtained from our DFT studies with that of the experiment.

| Property | GGA | GGA+vdW | Expt |
|---|---|---|---|
| Lattice parameter (Å) | 8.916 | 9.367 | 9.34349(3) |

As Ru$_3$Sn$_7$ is known to possess a complex electronic structure, we proceeded to the electronic structure calculations using the equilibrium structure that was established from our computation. The computed electronic density of states (DoS) is shown in Fig. 2. The total DoS (TDoS) at ambient conditions indicates that there is a continuous distribution of electronic states across the Fermi level ($E_F$), indicating the metallic nature of the system. The partial DoS (PDoS) shown in Fig. 2 shows that, just below the $E_F$, Ru states have a significant contribution to the total DoS rather than the Sn states with a substantial contribution from Ru-$d$ states. In the conduction band, the Sn states (both $p$ and $s$) contributions also have significant weight. A region with the lowest DoS has been also seen in the conduction band at the vicinity (~1.2 eV) of the $E_F$. The region below this, and above the $E_F$, is dominated by the contributions from Ru-$d$ and Sn-$p$ states with a small contribution from the Sn-$s$ state as well.

*Contact author: irshad.kariyattuparamb@elettra.eu
†Contact author: boby.joseph@elettra.eu

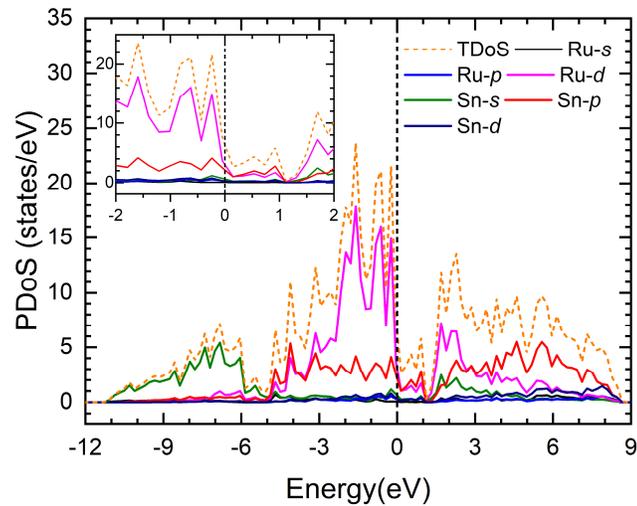

FIG 2. The spin-polarised total and partial electronic density of states (DoS) calculated for the ambient cubic structure of $Ru_3Sn_7$. Contributions from different orbitals of the Ru and Sn atoms are shown with solid lines having different colors, whereas the total DoS is shown with dotted lines. A significant contribution from Ru-$d$ states and Sn-$p$ states is evident. The inset figure shows the magnified view of partial DoS near the $E_F$. The total independent DoS contributions of the Ru and Sn atoms is provided in the Fig. S1 of the supplementary file. There is a significant DoS at the Fermi level ($E_F$) indicating the metallic nature of the system

To have further insights into the electronic states, the electronic band structures have been calculated as a function of pressure. The calculations have been performed with and without incorporating the spin-orbit coupling (SOC). The band structure at ambient conditions is shown in Fig. 3 (also in Fig. S2). The calculated band structure of $Ru_3Sn_7$ without the incorporation of the

*Contact author: irshad.kariyattuparamb@elettra.eu
†Contact author: boby.joseph@elettra.eu

SOC at ambient conditions indicates a Dirac-like electronic structure at various points along the Brillouin zone. A few of them lying near the $E_F$ are indicated in Fig. 3 and in Fig. S2. A Dirac cone-like feature at the $E_F$, where two bands (labelled as $B_1$ and $B_2$ in Fig. 3) meet each other is evident at the high-symmetry H point. Several other Dirac cone-like features are also seen in the vicinity of the H point (circled in Fig. 3). The comparison shown in Fig. 3 (and Fig. S2) indicates that SOC has a remarkable influence, which not only alters the energy level of various bands but also leads to the lifting of the band degeneracy at various regions. At several points along the Brillouin zone, the band crossings (without the SOC) open a gap with the incorporation of SOC. The conduction bands ($B_1$ & $B_2$) lying on the $E_F$ at the high symmetry H point are shifted apart in such a way that the $B_1$ shifts down to and the $B_2$ shifts away from the valence band. This separation of $B_1$ resulted in the formation of a new Dirac point just below the $E_F$ where the $B_1$ intersects one of the valence bands. As a consequence, a narrow gap opening of ~0.061 eV was evident at the $E_F$ through the high symmetry H point when the SOC was introduced. However, a finite direct energy gap is observed between the highest occupied valence band and the lowest unoccupied conduction band throughout the Brillouin zone, except at the H point.

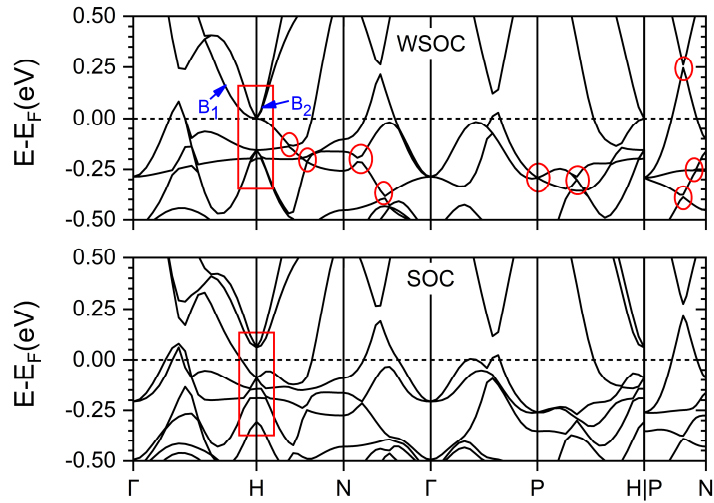

FIG 3. The band structure of $Ru_3Sn_7$ with and without the Spin-Orbit Coupling (SOC) indicating the presence of Dirac-like features (a few of them near the $E_F$ are marked with red circles). The region marked with the rectangle indicates the opening of a band at the H point due to the shift in $B_1$ and $B_2$ bands when SOC is incorporated

*Contact author: irshad.kariyattuparamb@elettra.eu

†Contact author: boby.joseph@elettra.eu

The Raman spectrum recorded at ambient pressure and room temperature from the $Ru_3Sn_7$ sample is shown in Fig. 4. We note that there were only few features (phonon modes) below 300 cm$^{-1}$ (see also Fig. S3). The Raman signal is found to be extremely weak, which may be due to the metallic nature of the sample. However, by careful investigations, we could reproducibly obtain a set of phonon modes below 300 cm$^{-1}$ which could be deconvoluted to six Raman modes: 121.9, 128.2, 157.9, 175.3, 196.4 and 228.4 cm$^{-1}$. The spectral deconvolution is also included in Fig. 4 (red solid lines). From our computational studies we could deduce the mode positions and characters of the expected 9 Raman modes, which are provided in Table III. Based on our computational results, we have assigned the experimentally observed modes which are $E_g(1)$, $T_{2g}(2)$, $T_{2g}(3)$, $E_g(2)$, $A_{1g}(1)/T_{2g}(4)$ and $A_{1g}(2)/E_g(3)$ respectively in the increasing order of the wavenumber. Out of these six modes, three of them are very weak. The predicted lowest wavenumber $T_{2g}(1)$ mode at 106 cm$^{-1}$ could not be detected in our experimental conditions due to the notch filter cut off blocking the Rayleigh line. A clear distinction between the $A_{1g}(1)$ with $T_{2g}(3)$ and $A_{1g}(2)$ with $E_g(3)$ could not be established indicating that they are either weak or degenerate within the experimental resolution. The calculated phonon DoS and the phonon dispersion are shown in Fig. S4. The phonon DoS calculations indicate that the vibrational modes below 150 cm$^{-1}$ are dominated by the contributions from the Sn atoms whereas both Sn and Ru atoms contribute equally above this. The vibrational configurations of the two strong $A_{1g}$ Raman modes are shown in the inset of Fig. 4. The $A_{1g}(1)$ mode involves the symmetric stretching vibrations of the Sn(2) atoms in the 16*f* position whereas the $A_{1g}(2)$ is due to the symmetric stretching vibrations of the Ru atoms in the 12*e* position. The remaining vibrational configurations obtained from the calculations are provided in Fig. S5 of the supplementary file.

*Contact author: irshad.kariyattuparamb@elettra.eu
†Contact author: boby.joseph@elettra.eu

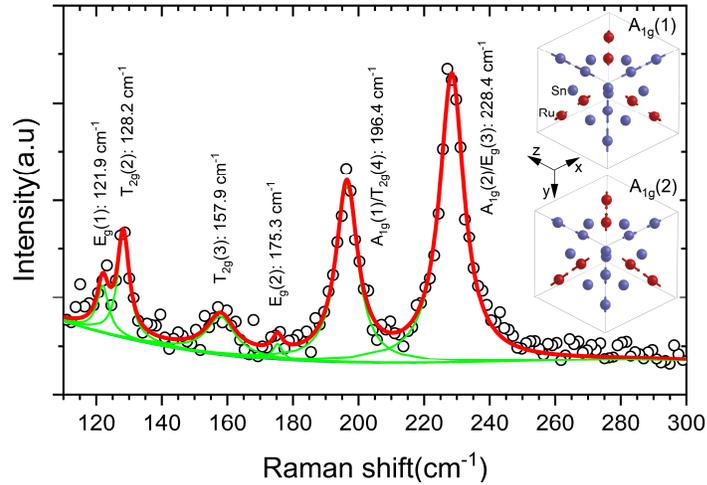

FIG 4 The Raman spectrum of $Ru_3Sn_7$ at ambient conditions (symbols) together with the results of the spectral deconvolution (red and green solid lines respectively indicating the cumulative fit and individual Lorentzian peaks). The vibrational configurations of the two high-intense $A_{1g}(1)$ and $A_{1g}(2)$ Raman modes are also indicated in the primitive cell of $Ru_3Sn_7$.

TABLE 3. The calculated and experimental Raman active modes along with the mode assignments. Only six out of the 9 Raman modes were resolved experimentally.

| Raman active mode | Theory (cm$^{-1}$) | Experiment (cm$^{-1}$) |
|---|---|---|
| $T_{2g}(1)$ | 105.94 | - |
| $E_g(1)$ | 122.23 | 121.9 |
| $T_{2g}(2)$ | 126.88 | 128.2 |
| $T_{2g}(3)$ | 156.82 | 157.9 |
| $E_g(2)$ | 174.69 | 175.3 |
| $A_{1g}(1)$ | 195.19 | 196.4 |
| $T_{2g}(4)$ | 195.95 | - |
| $A_{1g}(2)$ | 227.33 | 228.4 |
| $E_g(3)$ | 240.88 | - |

### B. High-pressure x-ray diffraction and Raman studies

In order to understand the structural evolution of $Ru_3Sn_7$, high-pressure (HP) x-ray powder diffraction (XRPD) data has been recorded till ~19 GPa. Apart from the shift in Bragg-peak positions towards the higher angles and the peak broadening, HP-XRPD data shown in Fig. 5(a) does not indicate any signature of structural phase transitions till the highest pressure studied, 19.2 GPa. It is also noted that a high degree of peak broadening is evident above 10 GPa. This is possibly due to the changing hydrostaticity conditions due to the solidification of the methanol ethanol mixture used as pressure-transmitting-medium in this study. Rietveld refinement has been carried out for all the diffraction patterns collected at higher pressures. The crystallite size and the atomic displacement parameters have been constrained to their ambient pressure values throughout the refinement process. The lattice parameter, microstrain parameter and the fractional atom positions were allowed to vary during the refinement. The results of the Rietveld refinement carried out at 6 and 18.2 GPa are shown in Fig. 5(b). There are no additional unindexed peaks thus, confirming the HP structural stability of the cubic phase of $Ru_3Sn_7$. We have attempted to compare and contrast the HP diffraction data of $Ru_3Sn_7$ with other isostructural compounds like $Ir_3Ge_7$, $Ir_3Sn_7$, $Mo_3Sb_7$, $Ni_3Ga_7$, $Pd_3In_7$, $Os_3Sn_7$, and $Re_3As_7$, that are reported in the literature [11,12]. Though these compounds are reported in ambient conditions, HP studies are not available for several of them. Among them, $Mo_3Sb_7$ has been studied extensively due to its superconducting nature at low temperatures and high pressures. Diffraction studies on $Mo_3Sb_7$ have indicated no structural phase transition till 13.6 GPa which is consistent with our present study [17]. Though a tetragonal structure was reported at a low temperature of 53 K for the $Mo_3Sb_7$, we could not see any signature of this structure for the $Ru_3Sn_7$ either. We could not find any other reports dealing with the high-pressure/ high-temperature structural studies of this family of systems.

*Contact author: irshad.kariyattuparamb@elettra.eu
†Contact author: boby.joseph@elettra.eu

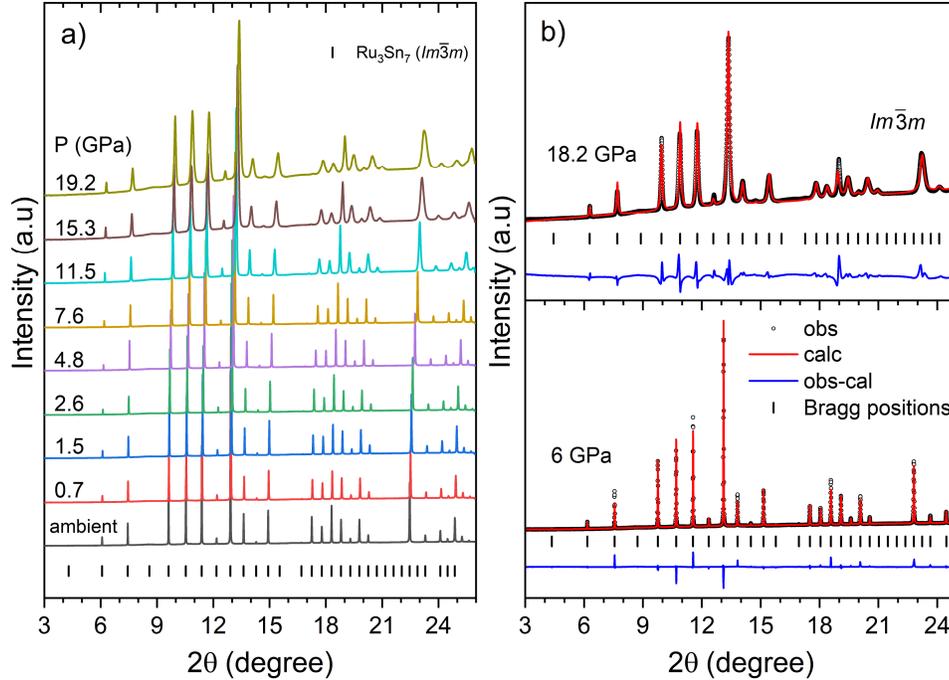

FIG 5. (a) The diffraction patterns of $Ru_3Sn_7$ at different pressures. Data are shifted in y for clarity in presentation. The tick marks at the bottom represent the Bragg positions of the cubic $Ru_3Sn_7$ (SG: $Im\bar{3}m$) at ambient conditions. The total number of Bragg peaks seems to remain similar at the highest pressure, thus ruling out the possibility of any structural phase transitions in this system till the highest pressure is studied. (b) The results of the Rietveld refinement analysis at 6 (bottom panel) and 18.2 GPa (top panel). Symbols are data and vertical lines are the Bragg peak positions. Model fit results using the SG: $Im\bar{3}m$ and the difference curves are shown respectively in red and blue curves.

*Contact author: irshad.kariyattuparamb@elettra.eu

†Contact author: boby.joseph@elettra.eu

Raman spectroscopy is a versatile tool to understand the structural, vibrational, and electronic changes in materials with extreme conditions. In many systems, the technique has been proven to be a promising tool for identifying the changes in the electronic structure under extreme conditions [35-40]. As there are no structural phase transitions observed in our study, Raman spectroscopy is expected to give further insights into the changes in electronic structure [37-40]. HP Raman spectroscopic measurements were carried out on the $Ru_3Sn_7$ sample to understand the phonon mode evolution with pressure. The Raman spectra collected at pressures up to 20.4 GPa are shown in Fig. 6(a). The weak $E_g(1)$, $T_{2g}(2)$, $T_{2g}(3)$, and $E_g(2)$ Raman modes could not be well resolved when the spectrum was acquired inside the DAC. This is mostly due to the absorption of the scattered spectral signal from these weak modes by the top diamond in our measurement geometry (backscattering). However, these peaks were feebly observed in some of the high-pressure patterns. Therefore, these peaks were not included in the convolution analysis. For the

*Contact author: irshad.kariyattuparamb@elettra.eu
†Contact author: boby.joseph@elettra.eu

analysed modes ($T_{2g}(2)$, $T_{2g}(3)$, $A_{1g}(1)$ and $A_{1g}(2)$), apart from the shift in frequency towards the higher wavenumber, the HP Raman data does not indicate any evidence of structural phase transitions till 20.4 GPa

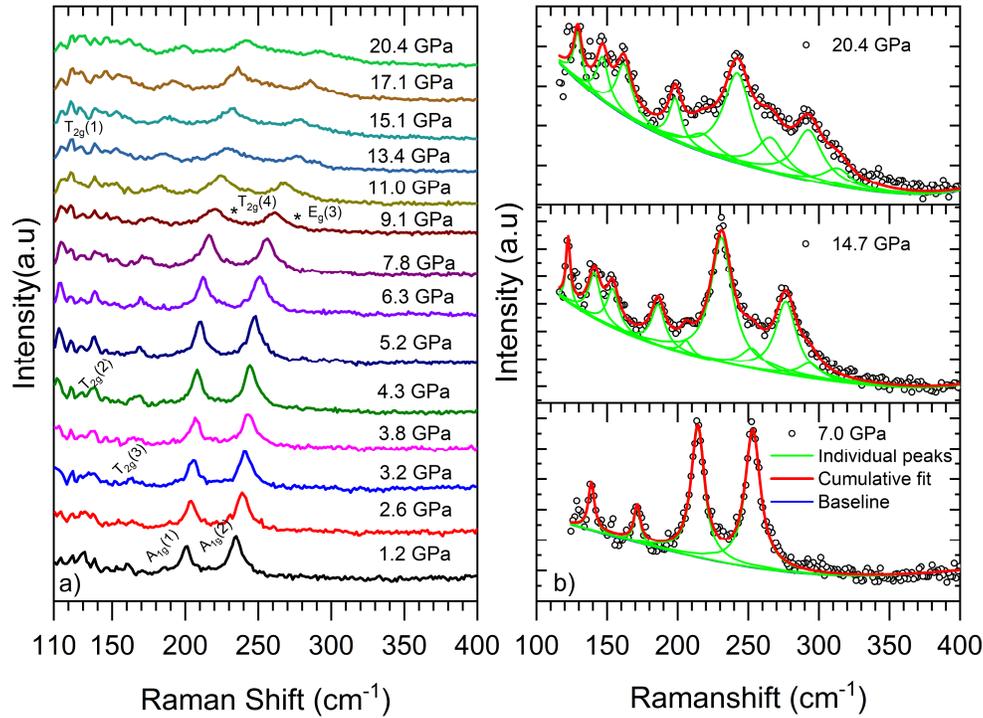

FIG 6. (a) Raman spectra of $Ru_3Sn_7$ at different pressures. The Raman spectra also seem to indicate the absence of any structural phase transitions. The asterisks indicate the $T_{2g}(4)$ and $E_g(3)$ modes that were resolved at higher pressures. At higher pressures, we could also observe the $T_{2g}(1)$ mode. (b) Examples of the Lorentzian spectra deconvolution of the Raman spectra at three representative pressures.

To have further quantitative information from the HP-Raman data, a careful spectral deconvolution using Lorenzian lineshapes were carried out. This analysis was done also keeping in mind the inputs from the computational study. In particular, the mode positions, FWHM, and the integrated intensity of different modes at each pressure have been obtained by fitting the data with the Lorentzian peak shape functions. Above 7.8 GPa, we constrained the peak width of all four modes to be similar. Three representative examples of the mode deconvolution by Lorentzian line shape fittings are shown in Fig. 6(b). Interestingly, with increasing pressure above 7.8 GPa, the Raman modes that were found to be degenerate at lower pressures seem to be resolved as shoulder peaks to their respective counterparts. These modes have been assigned as $T_{2g}(4)$ and $E_g(3)$ respectively in the increasing order of their frequency (indicated with *marks in Fig. 6(a)). To gain further knowledge about the existence of any structural phase transitions, we calculated the phonon dispersion at higher pressures. The dispersion spectra calculated at 3, 6, 8, and 12 GPa of pressures shown in Fig. S6. There are no imaginary frequencies in the calculated spectra, hence confirming that cubic structure of $Ru_3Sn_7$ is dynamically stable at all these pressures. This is in agreement with our inference from the HP-XRD. In consonance with our experimental observations, our phonon mode calculations could reproduce the lifting of the mode degeneracy in which the $A_{1g}(1)$ and $T_{2g}(4)$ modes were well resolved above 3 GPa (Fig. S7). In our experiment, we could not observe a complete separation of the shoulder peaks even at the highest pressure. However, their evolution indicates they show a more or less linear increase with increasing pressure. Further, a significant enhancement in the relative intensity of the $A_{1g}(1)/T_{2g}(4)$ mode with respect to the $A_{1g}(2)/E_g(3)$ mode is evident with increasing pressure. The intensity of the former mode gradually rises with the increase in pressure and then saturates. A qualitative analysis indicates that the intensity of both modes becomes nearly identical at around 7.8 GPa, above which these modes are no longer degenerate.

*Contact author: irshad.kariyattuparamb@elettra.eu
†Contact author: boby.joseph@elettra.eu

**C. Compressibility, equation of state, and phonon mode Grüneisen parameter (γ)**

To understand the compression behaviour, the lattice parameter and hence the volume of the unit cell has been obtained from the Rietveld refinement of the X-ray diffraction data at various pressures. The variation of the unit cell volume with pressure is shown in Fig. 7(a). A consistent reduction in the volume is evident with the increasing pressure indicating a smooth compressible nature of the cubic structure. A careful investigation shown in Fig. 7(a) indicate a clear deviation in the pressure evolution of the unit cell volume above the pressure of ~8 GPa. Such changes without any discontinuity in the pressure-volume data have been reported to be originated from the modifications in the electronic structure due to the electronic topological transitions (ETT) [37-40]. The possibility of the existence of pressure-induced ETT in $Ru_3Sn_7$ cannot be discarded as it retains topological states in its electronic structure [1,8]. Due to this divergence, the pressure-volume data has been fitted separately in these two regions using the 3$^{rd}$ order Birch-Murnaghan equation of state (EoS) employed in the Eosfit programme [41]. EoS fitting to the data below 8 GPa yields a bulk modulus value of 118.1(8) GPa with a $K'_0$ value 5.7(2). Assuming the derivative of the bulk modulus remains the same, EoS fitting to the data above 11 GPa indicates a higher bulk modulus value of 141 (2) GPa. To understand these changes, the compressibility curves obtained for other isostructural compounds were compared with the present study. The high-pressure behavior of $Mo_3Sb_7$ indicates no such changes in the compressibility, they obtained the bulk modulus as 111.5(8) GPa at 80 K and 112 (18) at 300 K [22,42]. An earlier diffraction measurement on $Mo_3Sb_7$ to a maximum pressure of 13 GPa indicated 126 (8) GPa as the bulk modulus which is similar to what we obtained for the $Ru_3Sn_7$ [17]. DFT calculations on $Ir_3Sn_7$ have reported 133 GPa as the bulk modulus, which is in the similar range of the values that we obtained for $Ru_3Sn_7$ [43]. The available reports indicate that the compressibility curve obtained from our study is in good agreement with the other isostructural compounds. However, the change in the compressibility behaviour that we observed here was neither observed in any other experimental studies nor predicted by the theoretical calculations. In fact, our own DFT calculations also do not reproduce this (See Fig. 7(a) inset). The calculation predicted a bulk modulus value of 106 GPa which is slight underestimation of the value obtained from the experiment. The difference on a small scale could be due to the difference in the calculated unit cell volume.

*Contact author: irshad.kariyattuparamb@elettra.eu
†Contact author: boby.joseph@elettra.eu

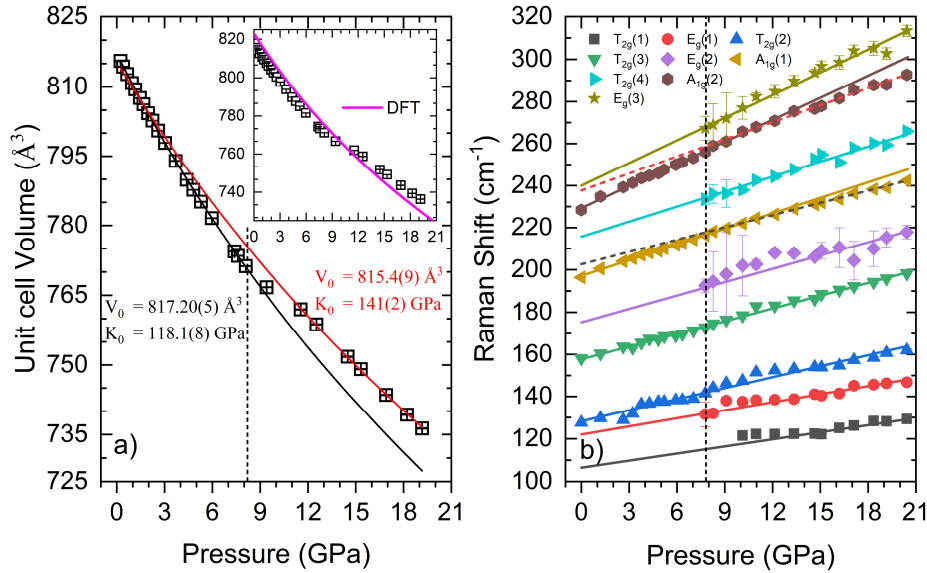

FIG 7. (a) The unit cell compressibility curve showing two different behaviours; above and below ~8 GPa. Symbols are experimental data. The black and red solid lines are the third-order BM equation of state fit. Inset figure compare the PV data obtained from the DFT calculations (solid magenta line) with experiments (symbols). (b) Evolution of the Raman modes as a function of pressure. The solid and dashed lines through the data points are the linear fit to the individual data sets. The vertical dashed line indicates the pressure at which the slope change occurs for the $A_{1g}$ (1) and $A_{1g}$ (2) modes.

*Contact author: irshad.kariyattuparamb@elettra.eu

†Contact author: boby.joseph@elettra.eu

Referring back to the inputs from the HP-Raman studies, we have noticeable changes in the Raman spectrum. A quantitative analysis can provide a deeper understanding of the underlying phenomena. The pressure evolution of the mode frequencies as obtained from the Lorentzian line shape deconvolution is shown in Fig. 7(b). All the mode frequencies are found to show hardening with the increase in pressure. However, a small but discernible slope change was observed for the $A_{1g}(1)$ and $A_{1g}(2)$ mode around 7.8 GPa which is also consistent with the deviation in the compressibility curve obtained from the HP-XRPD data analysis. As there are no structural transitions observed, such a change could be an indication of subtle changes in the electronic structure of the system [37-40]. In order to better describe the effect of pressure on these vibrational modes, the dimensionless quantity, Grüneisen parameter defined as $\gamma_i = (K_0/\omega_0)*(d\omega/dp)$ can be used. Here, $K_0$ and $\omega_0$ represent the isothermal bulk modulus and the mode frequency at ambient conditions, respectively. It is known that the microscopic lattice vibration and the macroscopic thermodynamic properties are closely connected by the $\gamma_i$. Due to the presence of a slope change, a linear fit to the pressure evolution of $A_{1g}(1)$ and $A_{1g}(2)$ modes has been carried out independently above and below 7.8 GPa. The $\gamma_i$ for these modes is calculated separately in the two regions using the zero pressure bulk modulus values obtained from the current study. For the modes, $T_{2g}(1)$ and $E_g(2)$, only the data above 13 GPa is considered to obtain the $\gamma_i$ values as we only see systematic behaviour above this pressure. The linear fit to these data points resulted in the $\omega_0$ values of 106(3) cm$^{-1}$ and 175(2) cm$^{-1}$ which are consistent with the computed values. Further, the $\omega_0$ values obtained from the linear fitting to the remaining modes are in good agreement with the experimental values. The calculated pressure evolution of the phonon frequencies shown in Fig.

*Contact author: irshad.kariyattuparamb@elettra.eu

†Contact author: boby.joseph@elettra.eu

TABLE 4. The calculated values of the $\omega_0$, d$\omega$/dp and the Grüneisen parameter ($\gamma$) for all the observed Raman modes obtained by linear fitting the pressure evolution of the modes. The $K_0$ values of 118.1(8) GPa and 141(2) GPa obtained from this study have been used to calculate $\gamma_i$ for pressures below and above 7.8 GPa, respectively.

| Mode | $\omega_0$ (cm$^{-1}$) | d$\omega$/dp | $\gamma_i$ |
|---|---|---|---|
| $T_{2g}(1)$ | 106(3) | 1.1(2) | 1.47(26) |
| $E_g(1)$ | 122(2) | 1.3(1) | 1.47(14) |
| $T_{2g}(2)$ | 128.2(5) | 1.74(6) | 1.61(06) |
| $T_{2g}(3)$ | 157.7(3) | 2.00(3) | 1.50(03) |
| $E_g(2)$ | 175.3(0) | 2.09(2) | 1.69(04) |
| $A_{1g}(1)$ [<7.8GPa] | 197.1(3) | 2.47(5) | 1.48(03) |
| $A_{1g}(1)$ [>7.8GPa] | 202.6(5) | 1.94(4) | 1.35(03) |
| $T_{2g}(4)$ | 215(3) | 2.38(2) | 1.56(11) |
| $A_{1g}(2)$ [<7.8GPa] | 229.3(4) | 3.50(9) | 1.80(05) |
| $A_{1g}(2)$ [>7.8GPa] | 238(1) | 2.70(8) | 1.61(06) |
| $E_g(3)$ | 240(2) | 3.6(1) | 2.11(08) |

*Contact author: irshad.kariyattuparamb@elettra.eu

†Contact author: boby.joseph@elettra.eu

S7 predicted a consistent uptrend for all these modes without any mode softening. In addition, we could also reproduce the slope change for the $A_{1g}(1)$ mode in our calculation. The experimental values of $\gamma_i$ obtained for all these modes are shown in Table IV. It is seen that the $\gamma_i$ of the $A_{1g}(2)$ mode shows a higher value compared with the $A_{1g}(1)$ mode in both regions. It is surprising that, despite the significant changes in $d\omega/dp$ and $K_0$ in both regions, the $\gamma_i$ remained nearly the same. It is to be understood that, the change in $d\omega/dp$ for these modes above 7.8 GPa is compensated by the changes in bulk modulus ($K_0$) that keeps $\gamma_i$ nearly constant throughout the pressure range studied.

### D. Electron-Phonon Interaction

The changes in the phonon dispersion relation and the anharmonic force constants of materials are expected to be reflected in the full width at half maximum of the Raman modes (FWHM) [44,45]. Its inverse relationship with phonon lifetime is invaluable in understanding the isostructural electronic transitions in crystalline systems [46]. In Fig. 8(a), the pressure-dependent variation in the FWHM of the two high-intense Raman modes $A_{1g}(1)$ and $A_{1g}(2)$ is shown. The FWHM of both the modes are independent of pressure till 5.9 GPa. A monotonous and consistent rise in the FWHM of these modes is seen thereafter. Such anomalous behaviour has been seen in many topological materials [46]. This anomalous rise in linewidth around 5.9 GPa can be attributed to the enhancement of the electron-phonon coupling (EPC) in $Ru_3Sn_7$ [36,46]. As the linewidth of both modes follows a similar increasing trend, a strong EPC is expected at higher pressures. In Fig. 8(b), the experimental intensity of the $A_{1g}(1)$ mode relative to that of $A_{1g}(2)$ mode is shown. The relative intensity of $A_{1g}(1)$ mode increases gradually till 3 GPa followed by a rapid increase till ~14 GPa and, it saturates at further higher pressures. As there are no structural phase transitions, such an increase in the intensity can be attributed to the enhancement in the electron-phonon scattering process at higher pressures. Further insights about the electron-phonon interactions can be obtained from the Eliashberg spectral function ($\alpha^2 F(\omega)$), which describes the strength of EPC as a function of phonon frequency[47]. However, computing $\alpha^2 F(\omega)$ for materials with large number of atoms in their primitive cell is tedious and sometimes computationally prohibitive. Since, the Eliashberg spectral function, $\alpha^2 F(\omega)$, is proportional to the phonon and electron (near the $E_F$) DoS, tracking their changes with pressure will provide a qualitative information about the $\alpha^2 F(\omega)$, and then the emergence and strength of EPC [47-50]. In Fig. 8(c), we show the pressure evolution of the total phonon DoS at 0 and 3 GPa. Additionally, the total and atom/site decomposed phonon DoS at 6, 8 and 12 GPa of pressures are shown in Fig. S8. The total phonon DoS clearly indicate a few narrow peaks with minimal spread in the 3-6 GPa of pressure. In addition, these peaks are higher in magnitude compared to the DoS obtained at 0 GPa resulting in the enhancement of the available number of vibrational modes of specific frequency to which the electrons can interact with. The partial phonon DoS show that, this enhancement is majorly due to the Ru atoms at 12*e* site and Sn atoms at 12*d* site. Nevertheless, it is evident that there is an overall increment in the phonon DoS with the increase of pressure from 0 to 6 GPa. In this case, a stronger electron-phonon interaction is expected. In order to understand the pressure effect on the electronic DoS, DFT calculations have been extended to four different pressures, 3, 6, 8 and 12 GPa. As representatives, the total electronic DoS obtained at 0 and 3 GPa is shown in Fig. 8(d). The electronic DoS for the remaining pressures are shown in Fig. S9. It is seen that, the total electronic DoS shows noticeable changes with pressure. At the $E_F$ and its proximity, the total DoS calculated at 3, 6 GPa shows an upsurge compared with the one calculated at the ambient pressure. It is to be noted that, we have already seen an increased number of phonon states at these pressures. Since there are no structural phase transitions, the increased number of electronic DoS together with an increased number of phonon DoS, broadening in the Raman line width and the enhancement of the Raman mode intensity signifies the onset of the electron-phonon scattering process, resulting in a stronger EPC at these pressures.

*Contact author: irshad.kariyattuparamb@elettra.eu
†Contact author: boby.joseph@elettra.eu

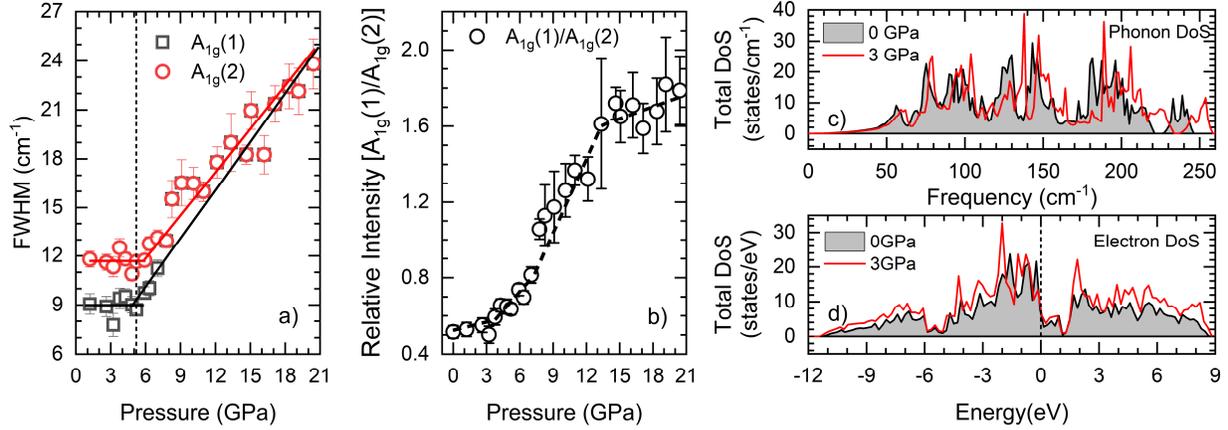

FIG 8. a) Pressure-dependence of the FWHM of $A_{1g}(1)$ and $A_{1g}(2)$ mode showing a sudden rise above 5.9 GPa. The onset pressure at which the rise in FWHM occurs is indicated with a vertical black dotted line. The solid lines through the data points are a guide to the eye. b) the variation in the relative intensity of $A_{1g}(1)$ mode to the $A_{1g}(2)$ mode with pressure. The dashed lines through the data points are guide to the eye. The total phonon (c) and electron (d) DoS obtained at 0 and 3 GPa of pressure, revealing the enhancement of DoS at 3 GPa.

*Contact author: irshad.kariyattuparamb@elettra.eu

†Contact author: boby.joseph@elettra.eu

### E. Electronic structure at higher pressures

The enhancement in the electronic DoS within the pressure range of 3-6 GPa has already been established in the earlier sections, which illustrate the apparent changes in the electronic properties of $Ru_3Sn_7$ at higher pressures. In addition, a shift of the total DoS towards the conduction band with increasing pressure just above the $E_F$ is also evident (Fig. S9). As a result, the total bandwidth (sum of valence and conduction band) increased from 20.127 eV at 0 GPa to 20.407, 20.835, 20.850 and 21.357 eV at 3, 6, 8 and 12 GPa, respectively. This increase in the bandwidth is a signature of the increasing delocalization of the electronic orbitals in $Ru_3Sn_7$. At pressures of 6 GPa and above, DFT calculations reveal a discrete distribution of electronic states above the Fermi level in the conduction band. This is in contrast with the characteristics of the total DoS obtained at ambient pressure in which a continuous distribution of the total DoS is observed across the conduction band.

*Contact author: irshad.kariyattuparamb@elettra.eu

†Contact author: boby.joseph@elettra.eu

To better understand the pressure effects and the origin of EPC, the orbital decomposed band structure of $Ru_3Sn_7$ is calculated at 0, 3, 6, 8 and 12 GPa, and the two representative band structure at 0 GPa and 12 GPa is shown in Fig. 9(a). A comparison with and without the SOC in Fig. S10 and S11 indicates that the SOC is strong even at higher pressures. The bands near the $E_F$ are predominantly contributed by the Ru-*d* and Sn-*p* states with a minor contribution from the Sn-*s* orbital, corroborating with DoS data. Upon increasing the pressure, the Ru-*d* band that was crossing the $E_F$ along the N→Γ region shifted the crossing point to the high symmetry N point (indicated as R1 in Fig. 9(a)). At 6 GPa, the hybridization of *spd* orbitals near the $E_F$ in the conduction band indicated as R2 in Fig. 9(a) vanishes retaining the *sd* overlap intact. With further increasing pressure, these bands shift away from each other considerably and eventually create a gap in the conduction band as also seen in the DoS. A significant observation is the band-crossing phenomenon as viewed in Fig. 9(b) (also indicated as R3 in Fig. 9(a)). In connection with this, several changes were observed near the $E_F$ at the high symmetry H point. As pressure increases, the Dirac cone-like feature observed in the conduction band close to the vicinity of the $E_F$ penetrates the valence band through the high symmetry H point at 8 GPa. At further higher pressure, the Sn-*p* orbital is pushed down to the valance band leading towards the closing of the band gap along this direction. Besides this, the $B_1$ band just below the $E_F$ show a gradual shift from predominantly Sn-*p* to Ru-*d* type character as seen in Fig. 9(b), suggesting a pressure induced hybridization of these orbitals. Moreover, the Sn-*p* band ($B_2$) from the conduction region is seen to overlap with the Ru-*d* orbitals in the valence band at higher pressures. In association with this, multiple numbers of Ru-*d* orbitals close to this energy show changes in their band curvature (see

*Contact author: irshad.kariyattuparamb@elettra.eu
†Contact author: boby.joseph@elettra.eu

Fig. 9(b)). As a consequence, a strong *dp* hybridization is set up between the Ru-*d* and Sn-*p* orbitals along the H point. It is to be noted that, the pressure at which the *dp* hybridization occurred is in agreement with the experimental findings in which a clear reduction in the unit cell compressibility and enhancement in the EPC was observed. Thus, it can be concluded that this pressure induced *dp* hybridization at the $E_F$ is responsible for the reduction in the compressibility and strengthening of EPC. These interactions are interconnected and play a crucial role in dictating the material properties.

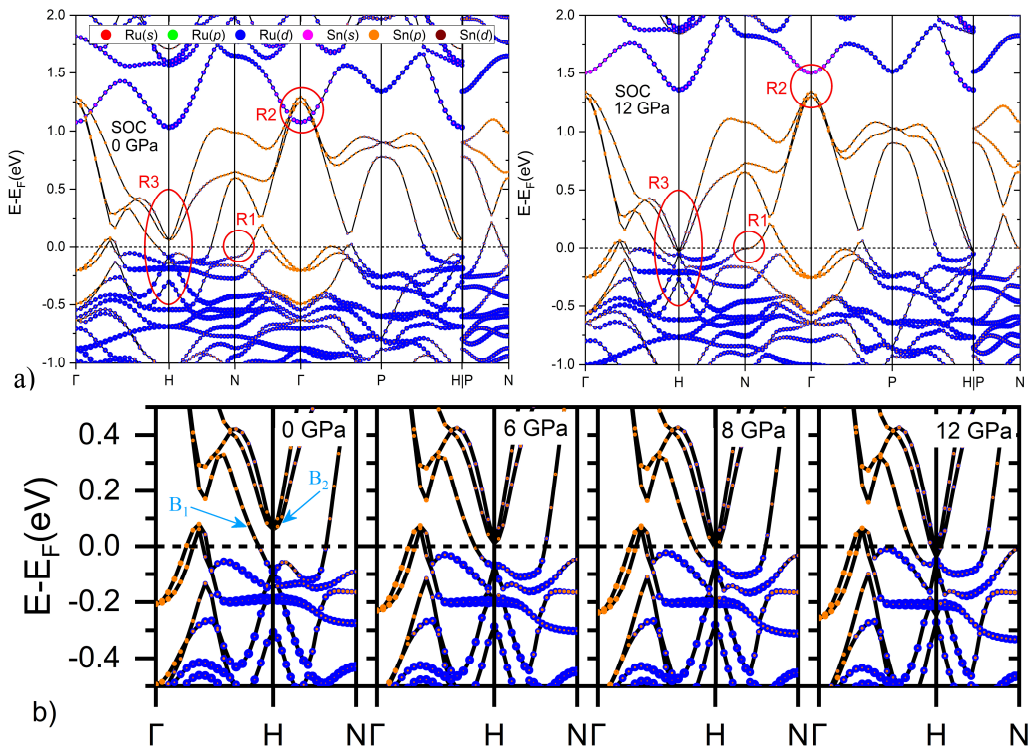

FIG 9. a) The orbital decomposed band structure calculated at 0 GPa and 12 GPa (top panel), the regions where major changes in the band structure occur with increasing pressure are indicated by circles (R1 and R2) or ovals (R3). For the explanation of different regions refer to the text. b) The magnified view of the band crossing through the $E_F$ at the H point. These results are with the inclusion of the SOC. At the H point, there is a clear evolution of *dp* hybridization with pressure.

*Contact author: irshad.kariyattuparamb@elettra.eu

†Contact author: boby.joseph@elettra.eu

## IV. CONCLUSIONS

High pressure (HP) X-ray powder diffraction (XRPD), Raman spectroscopy, and *ab initio* density functional theory (DFT) calculations have been carried out for the intermetallic compound $Ru_3Sn_7$ up to ~20 GPa. The HP-XRPD indicates no structural phase transitions up to 19 GPa. However, change in the compressibility was evident above 8 GPa leading to an increase in bulk modulus in this region. The lattice dynamic calculations revealed the presence of 9 Raman active modes in $Ru_3Sn_7$, with six observed under ambient conditions and the remaining three peaks resolvable at higher pressures. The Raman spectra indicate several changes at higher pressures. An enhancement in the intensity of the $A_{1g}(1)$ mode at ~3 GPa is followed by an anomalous increase in the pressure evolution of FWHM of the Raman modes above 5.2 GPa. These changes in the Raman spectrum together with the increased phonon and electron DoS around 3 GPa indicate an enhancement of the electron-phonon interaction. In the absence of any structural phase transitions, these changes are considered to be originating from the changes in the electronic structure. In support, the band-structure calculations substantiate that the observed discontinuities in the pressure dependence of the phonon modes and the enhancement in the electron-phonon coupling are related to the *dp* hybridisation along the high symmetry point of the Brillouin zone.

## ACKNOWLEDGMENTS

K. A. I acknowledge the Indian Institute of Science, Bengaluru, India and International Centre for Theoretical Physics, Trieste, Italy for the IISc/ICTP fellowship. K. A. I also thank Frederico Gil Alabarse of Xpress beamline of Elettra Sinchrotrone, Trieste for his scientific advices and guidance. Technical support received from Gianmario Skerlj, Bryan Ulivi and Andrea Stolfa of Xpress beamline is greatly appreciated.

*Contact author: irshad.kariyattuparamb@elettra.eu

†Contact author: boby.joseph@elettra.eu

*Contact author: irshad.kariyattuparamb@elettra.eu

†Contact author: boby.joseph@elettra.eu

*Contact author: irshad.kariyattuparamb@elettra.eu

†Contact author: boby.joseph@elettra.eu

*Contact author: irshad.kariyattuparamb@elettra.eu

†Contact author: boby.joseph@elettra.eu

*Contact author: irshad.kariyattuparamb@elettra.eu

†Contact author: boby.joseph@elettra.eu